# 3-D Magnetotelluric Investigations for geothermal exploration in Martinique (Lesser Antilles). Characteristic Deep Resistivity Structures, and Shallow Resistivity Distribution Matching Heliborne TEM Results

Nicolas Coppo[1], Jean-Michel Baltassat[1], Jean-François Girard[1], Pierre Wawrzyniak[1], Sophie Hautot[2], Pascal Tarits[3], Thomas Jacob[1], Guillaume Martelet[1], Francis Mathieu[1], Alain Gadalia[1], Vincent Bouchot[1] and Hervé Traineau[4].

[1]Bureau de Recherches Géologiques et Minières, 3 av. Claude Guillemin, BP 36009, 45060 Orléans Cedex 2, France.

[2]IMAGIR, Technopôle Brest-Iroise, Institut Universitaire Européen de la Mer, Site de la Pointe du Diable, Place Nicolas Copernic, 29280 Plouzané, France.

[3]Université de Bretagne Occidentale, Technopôle Brest-Iroise, Institut Universitaire Européen de la Mer, Site de la Pointe du Diable, Place Nicolas Copernic, 29280 Plouzané, France.

[4]CFG Services, 3 av. Claude Guillemin, 45100 Orléans, France

n.coppo@brgm.fr, jm.baltassat@brgm.fr, a.gadalia@brgm.fr, jf.girard@brgm.fr, Sophie.Hautot@imagir.eu, t.jacob@brgm.fr, G.martelet@brgm.fr, f.mathieu@brgm.fr, tarits@univ-brest.fr, h.traineau@cfg.brgm.fr, p.wawrzyniak@brgm.fr

**ABSTRACT**

Within the framework of a global French program oriented towards the development of renewable energies, Martinique Island (Lesser Antilles, France) has been extensively investigated (from 2012 to 2013) through an integrated multi-methods approach, with the aim to define precisely the potential geothermal ressources, previously highlighted (Sanjuan et al., 2003). Amongst the common investigation methods deployed, we carried out three magnetotelluric (MT) surveys located above three of the most promising geothermal fields of Martinique, namely the Anses d'Arlet, the Montagne Pelée and the Pitons du Carbet prospects. A total of about 100 MT stations were acquired showing single or multi-dimensional behaviors and static shift effects. After processing data with remote reference, 3-D MT inversions of the four complex elements of MT impedance tensor without pre-static-shift correction, have been performed for each sector, providing three 3-D resistivity models down to about 12 to 30 km depth. The sea coast effect has been taken into account in the 3-D inversion through generation of a 3-D resistivity model including the bathymetry around Martinique from the coast up to a distance of 200 km. The forward response of the model is used to calculate coast effect coefficients that are applied to the calculated MT response during the 3-D inversion process for comparison with the observed data. 3-D resistivity models of each sector, which are inherited from different geological history, show 3-D resistivity distribution and specificities related to its volcanological history. In particular, the geothermal field related to the Montagne Pelée strato-volcano, is characterized by a quasi ubiquitous conductive layer and quite monotonic typical resistivity distribution making interpretation difficult in terms of geothermal targets. At the opposite, the resistivity distribution of Anse d'Arlet area is radically different and geothermal target is thought to be connected to a not so deep resistive intrusion elongated along a main structural axis. Beside these interesting deep structures, we demonstrate, after analyzing the results of the recent heliborne TEM survey covering the whole Martinique, that surface resistivity distribution obtained from 3-D inversion reproduce faithfully the resistivity distribution observed by TEM. In spite of a very different sampling scale, this comparison illustrates the ability of 3-D MT inversion to take into account and reproduce static shift effects in the sub-surface resistivity distribution.

## 1. INTRODUCTION

The geothermal potential of Martinique Island (Lesser Antilles, France) has been extensively investigated since more than 50 years. The first exploration started in the 60' in the Plaine du Lamentin (south of Fort-de-France) and has been extended in this area in the 80's and orientated towards the southeastern flank of the Montagne Pelée. Then, the southwestern flank of the Montagne Pelée and the Anses d'Arlet) were identified of potential interest and explored in the 2000's. Finally, three exploration wells were drilled in the Plaine du Lamentin in 2001 and confirmed the presence of a medium temperature geothermal resource.

This project had for main objectives to first compile existing knowledge about the geothermal potential of Martinica Island, but also to proceed to adequate additional field exploration with the aim to complete and strengthen the existing geoscientific database and finally provide an up-to-date interpretation for the four selected areas (Anses d'Arlet, Montagne Pelée, Pitons du Carbet and Lamentin) with proposition of the most favorable zones for exploration wells. New geological (exploration of mineralogical clues related to magmatic and hydrothermal activity and structural context), hydrogeological (geochemistry and recharge assessment), geochemical (soil and water gas analysis) and geophysical data were acquired on these four sectors.

Beyond crucial geological, geochemical and hydrogeological data, geothermal exploration requires additional images of the physical properties of subsurface to understand and interpret properly the geothermal reservoir and its potentialities. This information can only be provided by adapted geophysical measurements above areas of interest. Because the amount of data collected during this two-years project is too important to be rigorously presented and described in a single communication, this paper mainly focuses on the geophysical results (new magnetotelluric and gravimetrical data, collected over the Anses d'Arlet and the Montagne Pelée province.

## 2. GEOLOGICAL AND GEOTHERMAL OVERVIEW

The Lesser Antilles island arc, which results from the westward subduction of the Atlantic Plate under the Carribbean Plate, has been active since the Early Eocene (Nagle et al., 1976). Arc evolution has been described as occurring in three stages, representing the old, intermediate and recent arcs, corresponding to Late Oligocene-Early Miocene, Middle Miocene and Late Miocene to recent ages, respectively (Briden et al., 1979; Westercamp, 1979), and having erupted from east to west. Due to the geological evolution of this subduction zone, Martinique is the northernmost island where the products of the three stages crop out from east to west without superimposition.

At the end of the Miocene, the locus of volcanism jumped to the north of the island producing the Morne Jacob shield volcano, whose oldest rocks are dated at 5.14+-0.07 Ma (Germa et al., 2010), making the onset of the recent arc activity in Martinique Island. From 2.3 Ma, Trois Ilets volcanism commenced in the southwest, coeval with continued activity at Morne Jacob volcano (Germa et al., 2011a). Trois Ilets Volcanism, Carbet Complex and Conil-Pelée Complex were then active contemporaneously until 345 ka (Germa et al., 2011a). Most recently, volcanic activity has been restricted to the northern Conil-Pelée Complex, whose last eruption occurred in 1929-1932. Lavas from the recent arc are tholeitic (Morne Jacob, early stage) to calc-alkaline, with basaltic-andesite to dacite compositions (Labanieh, 2009; Germa et al., 2010, 2011a). Each complex has its own geochemical signature, for major and trace elements, and for isotopic composition (Smith and Roobol, 1990; Labanieh, 2009; Germa et al., 2010, 2011a).

### 2.1 Anses d'Arlet geothermal field (see Gadalia et al., 2015, WGC)

Recent geological observations and analysis collected during this project have improved and specified the geological context of the Anses d'Arlet prospect (Traineau et al., 2013). In particular, this geological work provides a consistent framework to the magmatic and hydrothermal activity of the southwest of the Trois-Ilets Peninsula. Detailed mapping of the fossil alteration zones connected to their tectonic controls allowed to better constrain the geothermal reservoir context and its evolution up to now. A three steps continuous evolution is proposed in agreement with previous works by Germa et al. (2011) and Westercamp et al. (1990).

The initial stage is characterized by the development of a hydro-magmatic system thought to be related with the polyphased activity of the Roches Genty eruptive center (not dated). Many volcanic edifices grew up from north-west to south-est (Pointe Burgos-Le Diamant volcanic axis date between 1.5 and 0.35 Ma). Geological mixing between deep basaltic lavas and more superficial dacitic lavas with quartz suggests that magmatic differentiation occurred at relatively shallow depth (6 km from mineralogy). It is not obvious whether a single or multiple magmatic chambers fed all of these the volcanic edifices. The local tectonic setting (NW-SE and NE-SW structures) of the Trois-Ilets peninsula may have favored successive ascents of a deep basaltic magma during at least 1.1 Ma. These magma chamber(s) and associated intrusions would have been the heat source of an old acidic hydrothermal-magmatic system (early hydrothermalism) related to Roches Genty eruptive center.

The intermediate stage is characterized by the decline of the magmatic component and the evolution of the hydrothermal-magmatic acidic system toward a large high temperature neutral geothermal system. A clay cap develops together with a surface fumarolic activity along a corridor connecting the Anses d'Arlet and Petite Anse, reflecting the evidence of boiling related to thishigh temperature fossil system. Structural analysis demonstrates that faults reworked many time in stress tension creating favorable conditions for fluid circulation. This hydrothermalism can be described by three alteration products, spatially distinct: evidence of a kaolinite-alunite alteration (relief), the presence of sinter (silica deposits) and the existence of clay mineral with smectite (dominant) and trace of interstratified illite/smectite. During the last 0.5 Ma, the Morne Jacqueline experienced a lateral collapse that makes this fossil caprock outcropping together with intrusions.

The final stage represents the current geothermal activity expressed by the Eaux-Ferrées thermo-mineral springs in the Petite Anse bay where degassing and small travertine deposits can be observed. The small flow (0.03 l/s) of the springs associated with the fossil character, fumarolic and of high temperature, characteristic of the alunite-kaolinite alteration suggest a reduction both in extension and intensity of the surface manifestations. Deeper this could indicate a contraction (or a migration) of the HT geothermal system southwards.

### 2.2 Montagne Pelée geothermal field

Recent geological observations carried out during this project and reported by Traineau et al. (2013) completed the geological knowledge of the Montagne Pelée. Particularly, the pelean volcanic system, its formations, its magmatic heat source, its volcano-tectonic structures and the context of the surface hydrothermal manifestations disseminated on the southwestern flank of the stratovolcano. The Montagne Pelée is one of the most active volcano of the Antilles with 28 magmatic eruptions during the last 16 ka (Germa et al, 2011). Edifice growth and its plumbing system started about 127 ka ago (Germa et al, 2010; Boudon et al, 2013). Lavas geochemistry shows that the magmatic system nowadays feeding the Montagne Pelée is the same that fed Mont Conil in the past, both being located on the same NW-SE volcanic axis, the major tectonic direction of Martinica.

The main surface hydrothermal manifestations testifying a deeper geothermal activity are mainly located on the southwestern flank of the volcano. They are sulfuric springs of the uphill Rivière Claire (nowadays disappeared), fumaroles of the Etang Sec (also disappeared), hot springs from rivers Chaude Mitan and Picodo (still existing). The geochemical components of the two latters indicate the existence of another geothermal reservoir that could be separated from the main one. Hot spings and well located downhill on the coast are interpreted as a lateral extension of the river Chaude spring system. Other "cold" springs have a geochemistry indicating a deep component but very diluted.

No recent tectonic structure, except the three lateral collapses scars proposed by Le Friand et al. (2003) were identified on the southwestern flank. The first flank collapse occurred about 100 ka ago with a volume estimated to 25 km$^3$, the second (13 km$^3$), about 32 ka ago. Both of them considerably modeled the topography of the southwestern flank of the Montagne Pelée. We consider they had major implications on the geologic volcanic deposits pile accumulated around the volcano, causing deeper altered part of the edifice to reach the surface with a probable thinning of the caprock. Precise triggering factors of such catastrophic event are still a matter of debate but can be related to volcanic activity, caldera collapse, strong rainfall, climatic change (Coppo et al., 2009,

Quidelleur et al., 2008), seismic activity, fault activity. A third smaller (2 km$^3$) flank collapse occurred about 9 ky ago and had a reduced impact of the stratovolcano structure.

To the summit of the Montagne Pelée, two calderas and eruptive conduits have been identified by Westercamp and Traineau (1983). A larger old one has been under discussion since the evidence of lateral collapse.

## 3. MAGNETOTELLURIC METHOD

Different rocks, sediments, geological structures and fluids have a wide range of different electrical conductivities. Therefore, measuring electrical resistivity allows different materials and structures and fluids to be distinguished - sometimes identified - from one another and can improve knowledge about type of fluid content and its distribution, clay content, tectonic and geological structures.

The magnetotelluric method is a passive geophysical technique for imaging the electrical conductivity and structure of the Earth from the near surface down to several tens of kilometers (Vozoff, 1991). Its principle consists in simultaneous measurements and recordings of natural (internal) variations of electric and magnetic fields at the earth surface whose strength and geometry are governed by the sub-surface resistivity distribution. The primary external natural signal comes mainly from both the solar wind that induces pulsations in the magnetosphere and the lightning activity which both cause natural variations in the earth's magnetic field, inducing electric currents flowing under the Earth'surface. These naturally varying electromagnetic fields are measured over a wide range of magnetotelluric frequencies (from 10 kHz to 1000 s in our case). Low frequencies (< 1 Hz) are generated by the solar activity and highest (> 1 Hz) by worldwide thunderstorm activity. Combined, these natural phenomena create MT source signals over the entire frequency spectrum. Therefore, data quality is strongly dependent on external source activity and can be severely reduced - especially in the low frequency range – during phases of poor solar activity.

Magnetotelluric data comprises time series of the horizontal electric components (Ex and Ey measured in orthogonal directions) with their two associated magnetic components (Hx and Hy). When possible the vertical magnetic component (Hz) is also recorded. The ratio of the electric field to the magnetic field (E/H) at selected frequency, known as the "impendance tensor" embodies the full information about sub-surface conductivity structure. The ratio E/H is usually displayed as both apparent resistivity and phase as a function of frequency. Because of the "skin effect" phenomenon (explaining the attenuation of electromagnetic fields with depth), the depth of penetration of EM waves increases with lower frequency and higher resistivity. At most sites, collocated transient electromagnetic data were also acquired to correct the magnetotelluric data for the amplitude of galvanic distortion (Sternberg et al, 1988) caused by small-scale, shallow conductors.

A variety of techniques are available to assess the dimensionality of the causative structures from the impedance tensor, and, in the case of a two-dimensional (2D) Earth, determine the geoelectrical strike direction (Simpson 2005, Caldwell et al., 2004). In the present work, 3-D inversion has been directly applied to integrate tensors multi-dimensionality and sea-effect (see below).

### 3.1 Data acquisition

MT data have been acquired using Metronix MT stations (four ADU06 and three ADU07), MFS06e, MFS07 and MFS07e induction coils (for the magnetic field) and unpolarizable electrodes for electric field (Petiau, 2000). All MT stations were GPS synchronized. A remote reference has been permanently deployed in the Pitons du Carbet massif, in the center part of Martinique Island during the survey. Every day three new MT sites were occupied according to the survey design and objectives, and to local constraints (accessibility, authorization, proximity of noise sources). In order to increase data quality in the low frequency band (< 1Hz), our acquisition program has been extended to two days (three during the week-end). A total of 36.5 hours of low frequencies (sampling frequency = 256 Hz), 2 hours of intermediate frequency (sampling frequency = 2048 Hz) and 9 minutes of high frequencies (sampling frequency = 65536 Hz) were recorded. The remote reference was programmed to run on a daily basis in order to be precisely synchronized with the 6 MT stations deployed simultaneously on the field. Magnetic sensors were systematically buried at about 20-30 cm depth and electric and magnetic cables fixed or buried in the ground. A bentonite mud was prepared to insure good and long contact resistance of unpolarizable electrodes. Data quality revealed to be during better November than in December 2012 because of a lower solar activity during the last period as shown by comparison with geomagnetic observatories.

### 3.2 Processing

Before processing, data format was first homogenized and data arranged in a database dedicated for processing. Due to the large amount of data collected, a daily quality check was performed to insure both acquisition parameter accuracy, correct synchronization with the remote reference and data quality. We then applied a robust processing with remote reference for frequencies above 0.1 Hz based on a robust estimation of the impedance tensor errors described in Wawrzyniak et al. (2012). This statistical approach, based on BIRRP code (Chave and Thompson, 2004), allows to improve impedance tensor estimates providing that time series are long enough. Consecutively, low frequencies were processed using BIRRP. MT tensors and interpretational quantities were then plotted as function of frequency for analysis.

### 3.2 3-D MT inversion

Because of the complex volcanic island context, merging proximity of the sea with complex shoreline morphology, combined conical structure (with radial and circular geological objects) of a volcanic edifice and the linear main constitutive tectonic structures, 3-D inversion has been run with sea effect taken into account, to avoid potential distortion of the MT tensors. This task has been performed using the minim3D code (Hautot et al., 2000, 2007). A first solution to solve this effect is to consider the shoreline and the bathymetry around Martinique island as model parameters. However, a compromise between model resolution and time computation had to be chosen for the inversion. Furthermore, taking a coarse bathymetry and shoreline may result in numerical instabilities or imprecision. Therefore, correction coefficients for sea effects have been computed prior to the inversion, using digital elevation model of Martinique and existing detailed bathymetry (NOAA). Electric field **E** and magnetic field **B** computed at the surface of a "normal" medium (without sea, respectively $E_n$, $B_n$), i.e. our 3-D resistivity model during inversion, are related to the "true" fields (with sea, respectively $E_t$, $B_t$) by distortion tensors **K** and **R** using:

$$E_t = K^{-1} E_n \qquad (1)$$

$$B_t = R^{-1} B_n \qquad (2)$$

K and R distortion matrices have been computed using bathymetry until a distance greater than induction length of the maximum period acquired (1000 s), i.e. a distance about 200 km around each prospected area. The maximum resolution of the model in the investigation zone is 350 m. Then, during 3-D inversion process, at each iteration during forward model computation, E and B fields are corrected of the distortion factor before comparison to real data. The eight components of the impedance tensor are tried to be fitted during the 3-D inversion process. Tipper has not been inverted.

## 4. RESULTS & INTERPRETATION OF ANSES D'ARLET GEOTHERMAL FIELD

32 magnetotelluric soundings distributed in the area of interest (the uppermost three by helicopter owing to the rugged topography) were measured during the end of November 2012. After processing (see section 3.2), data analysis showed that most of the data exhibit a 1-D behavior above 1 to 10 Hz, while the longest periods reflect more complicated structures with sometimes diagonal components one order of magnitude lower than off-diagonal ones.

We also observed a lack of coherency around 10 s which affected data quality in this dead band (Figure 1). Although surficial resistivity is generally below 50 $\Omega$m in the investigated area, static shift effect was also observed at some stations. Before 3-D inversion, model parametrization has been designed to optimize the number of parameters versus the number of parameters, especially cells are merged together as a function of depth to reduce model parameters. A model of 18x16x12 cells (in the north, east and to depth direction, respectively is then adjusted to the grid data. The smaller cell of the model is 425x425m wide in the center of the model and the largest measures 2000 m at the boundary.

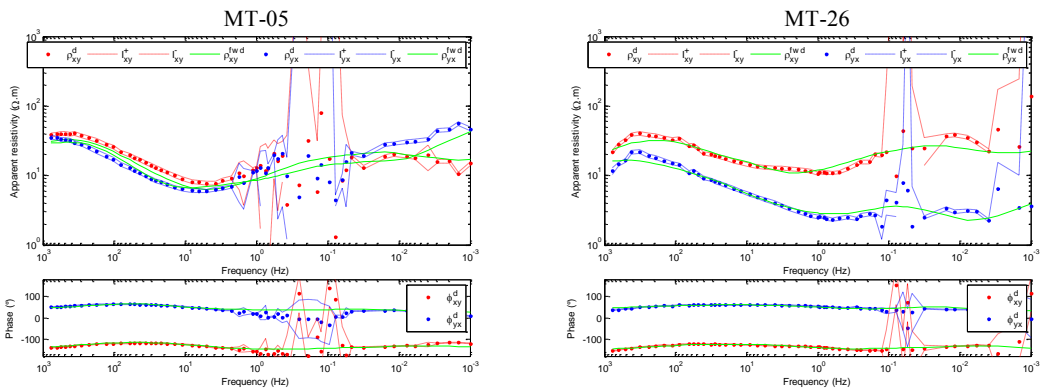

**Figure 1: MT soundings 05 (left) and 26 (right) showing data (off-diagonal Zxy and Zyx components in red and blue, respectively) and model responses in green. Upper plots display resistivity and lower plots phase for the same components.**

Figure 2 shows four layers of the resistivity model of the Anses d'Arlet obtained through 3-D inversion (very surficial resistivity distribution is presented on Figure 7 together with TEM data for comparison). Resistivity distribution of layer 5 (351-621 m depth) displays the emergence of a resistive costal body, globally orientated NNO-SSE, more evidenced around Morne Champagne (MCh, NNO). Inland, conductive areas mostly trail off excepted below Morne Larcher where a thickening is observed. Note the persistence of the conductive body centered on the hydrothermally altered zone between the Anses d'Arlet and Petite Anse (MT soundings 22 and 23), and elongated in the main tectonic direction of Martinique. Further north, a conductive body appears below Morne Réduit, and could be interpreted as for Morne Larcher to the altered core of the volcano.

Resistivity distribution of layer 6 (621-1081 m depth) illustrates the main geoelectric structure of the investicated area. The coastal resistive body identified on layer 5 is better defined and extends form Petite Anse (south of Morne Jacqueline) towards the Morne Réduit flanks. Its wide extent, 6 km along the NNO-SSE axis and 2.5 km perpendicularly suggests the presence of a relatively massive body. Only two conductive spots remain, one below Morne Larcher and the other below Morne La Plaine (northwards) delimited by two faults approximately orientated NE-SO. Although no MT sounding was recorded at the Anses d'Arlet, we observe a more conductive area always existing close to the hydrothermally altered zone, and disconnecting clearly the Morne Jacqueline. Note the southern well-defined boundary of the resistive body which concentrates most of the Petite-Anse springs having geothermal characteristics.

Resistivity distribution of layer 7 (1081-1851 m depth) shows the continuity with the overlaying layer 6. Northwestwards a homogeneous zone of 10-20 $\Omega$m appears and contrasts with the massive resistive body (10-200 $\Omega$m) along a N-S axis. The latter shows two poles because of the absence of MT data in the Anses d'Arlet area to constraint more precisely its extent. The conductive body of the Morne Larcher is still present but they are no more evidence between the hydrothermally altered zone and resistivity, suggesting that it is not so deeply rooted or that size may not be big enough to be highlighted.-

The last layer 8 (1851-3051 m depth) is globally similar to layer 7. Because of increasing depth, resolution decreases, but structuration remains the same. The main axis, now N-S is generated by the association of mesh cells at depth. We note a resistivity increase for the massive body and the terrain of intermediate resistivity.

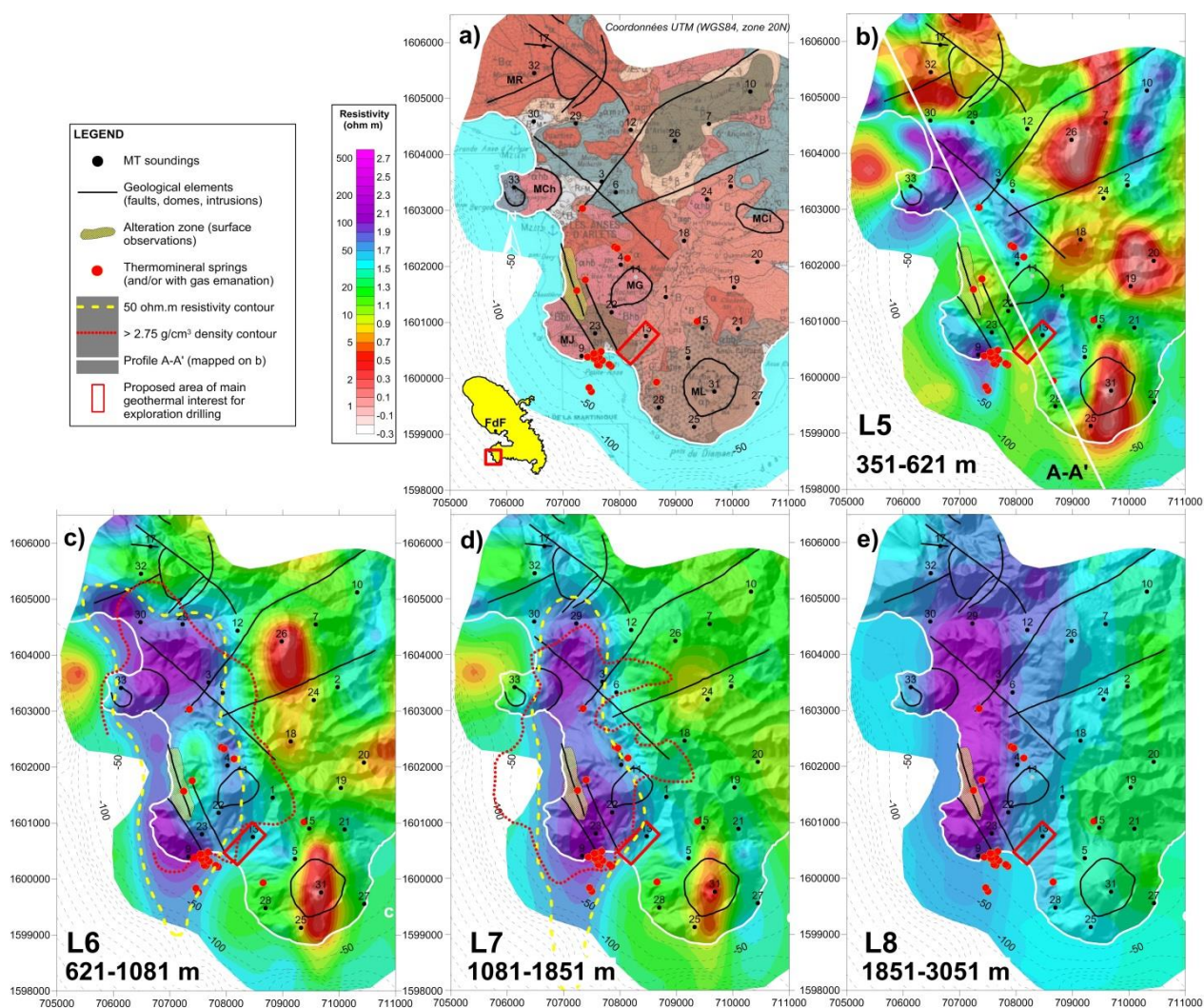

**Figure 2 : a) Location map of the investigated area on geological background (Westercamp et al. 1900) and resistivity models at 351 to 3051m deep(b, c, d, e). Superimposed elements are indicated in the legend.**

Geological observations around Morne Jacqueline, together with geochemical and previous geophysical results lead to an integrated interpretation of the area, illustrated on figure 3. The coast N-S resistive body is interpreted as a shallow intrusion deeply buried, which could be related to the observed superficial dioritic dikes in the Eaux Ferrées sector of Petite Anse (Traineau et al., 2013, Gadalia et al., 2014). This resistive anomaly (> 50 Ωm) is strongly correlated with the density model (contour with density > 2.75 g/cm3, Figure 3). We suggest that it could act as the heat source for the current geothermal system characterized by a probable decreasing activity (Gadalia et al, 2014). This interpretation does not discard a potential heat source related to the Morne Larcher (346 ka). Conductive spots are associated for some of them to the internal altered structure of volcanic edifices that could have developed a local hydrothermal system during eruptions phase. Following the classical model of Johnston et al (1992), the uppermost conductive layer (thicker in the southern part of the area) in interpreted as a cap rock of the geothermal system The geophysical synthesis proposes the area of Petite Anse to be the most promising for exploration (red square of circle on Figures 2 and 3).

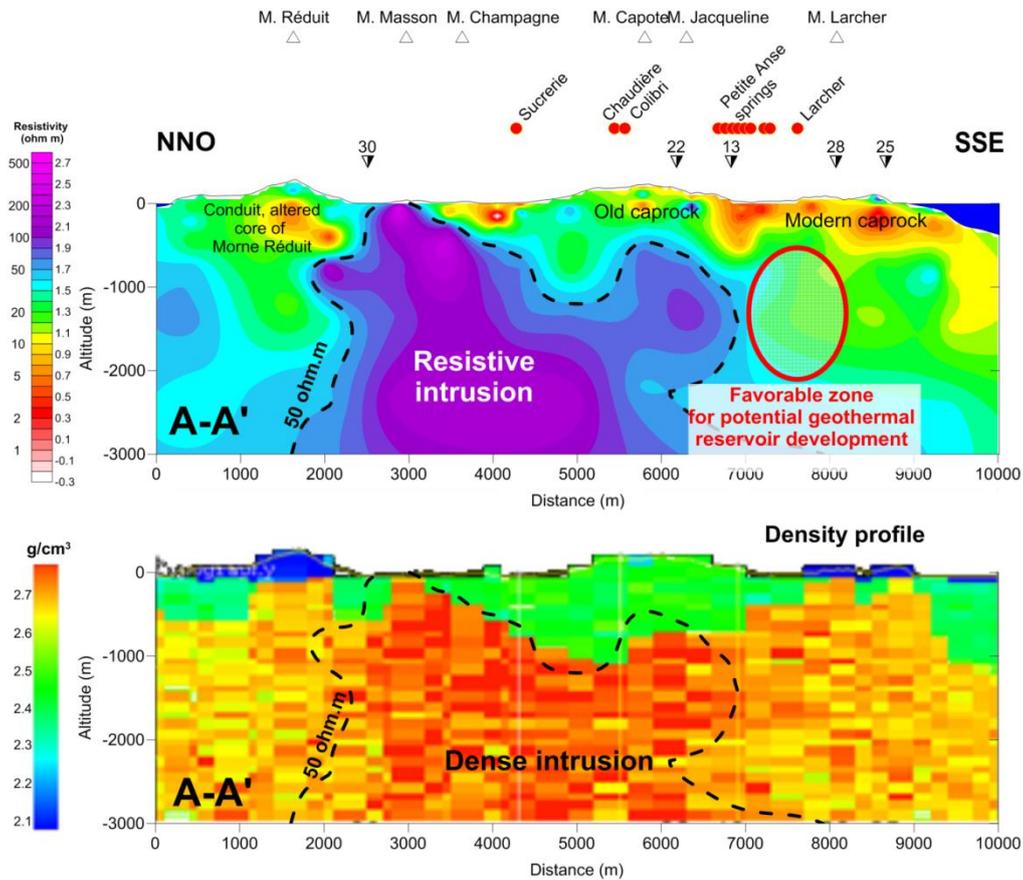

**Figure 3 : 2-D profiles crossing the area of main interest for further exploration drilling. Resistivity profile (above) is extracted from the 3-D resistivity model and the density profile (below) from the 3-D density model.**

## 5. RESULTS & INTERPRETATION OF MONTAGNE PELEE GEOTHERMAL FIELD

53 magnetotelluric sites were occupied (ten by helicopter owing to rugged topography) and distributed in the area of interest during December 2012 and January 2013. The southwestern flank of the Montagne Pelée has been extensively investigated while a tenth of MT soundings were acquired on the eastern flank to properly image potential interesting structures located below the summit.

Processing (see section 3.2) and data analysis showed that most of the data exhibit a 1-D behavior above 10 to 1 Hz, while the longest periods reflect more complicated structures. Signal induced by solar activity in the longest periods was significantly lower during December 2012 resulting in noisy components of the impedance tensor in the MT dead band (3 to 32 seconds) affecting resolution of the model at depth. Figure 4 shows typical results collected during the survey. A 3-D inversion has been run (Hautot et al., 2000, 2007) taking into account coast and sea effects. In order to resolve at best the complex structure of Montagne Pelée, model parametrization has been designed to optimize the number of data versus the number of parameters, especially cells are merged together as a function of depth to reduce model parameters. A model of 22x23x12 squared cells (in the north, east and to depth direction, respectively) is then adjusted to the grid data. The smaller cell is 500m wide in the center of the model and the largest measures 750 m at the boundary.

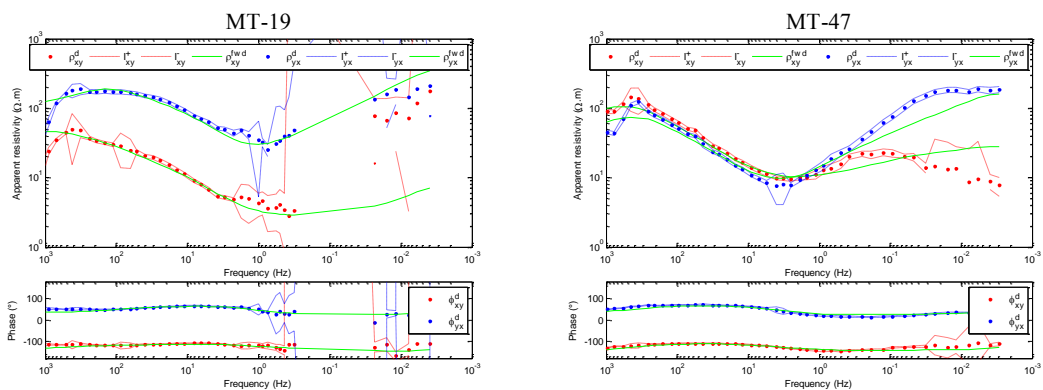

**Figure 4 : MT soundings 19 (left) and 47 (right) showing data (off-diagonal components in red and blue) and model responses in green. Upper plots display the resistivity and lower plots the phase for the same components.**

MT investigations on the Montagne Pelée reveal a first surficial resistive (20 < ρ < 2000 Ωm) layer 200-500 m thick attributed, at least for a part, to volcanic deposits related to the last eruption cycles of the Montagne Pelée. This unit overlays a thick (500-1000m) and massive conductive (ρ < 15 Ωm) which follows approximately the topography. Few major discontinuities apparently affect this layer that could be related with the long term volcanic and tectonic history of the Montagne Pelée. The conductive layer is mainly attributed to older deeply altered volcanic series (pre-Pelée?). Its high conductivity indicates high clay content which is unfavorable (at large scale) to fluid transfer and may play a caprock role above potential geothermal system(s). Despite the masking effect of this layer, below the conductive layer, MT data reveal a third and thick more resistive layer that can be interpreted in some parts as the potential geothermal reservoir. Deeper, the 1 km thick resistive layer is poorly constrained by the data although refraction seismic data (re-processed from Eschenbrenner et al., 1980) show a good morphological correlation with it, resulting of increasing velocity.

Figure 5 presents the results of 10 layers extracted from the 3-D resistivity model by altitude to illustrate characteristic changing resistivity with depth. From 800 m down to 200, a typical sub-circular resistivity distribution with radially increasing resistivity from the center highlights the conductive core of the volcano. The central conductive (< 5 Ωm) structure is interpreted as altered core thought to be related with the development of hydrothermal alteration during and post eruptive phases. Going down (200 m), the conductive layer extends forming a bell-like geometry whose top follows approximately the topography of the edifice. The presence of the conductive layer at higher altitude on the northeastern part of the volcano is strongly related to the topography and its historical evolution. The three lateral collapses that affected the southwestern flank of the Montagne Pelée during the last 100 ka (Le Friant et al., 2003) deeply excavated the structure of the volcano, thought to be initially more homogeneous. At sea level, we observe the development of a conductive body NNW-SSE just to the south of the summit, rotating to NW-SE and growing 200m below sea level. Note the changes between 0 and 400 m below sea level where the conductive layer has moved southwestwards leaving the upper part of the southwestern flank and the eastern part of the volcano with intermediate resistivity (10-50 Ωm).

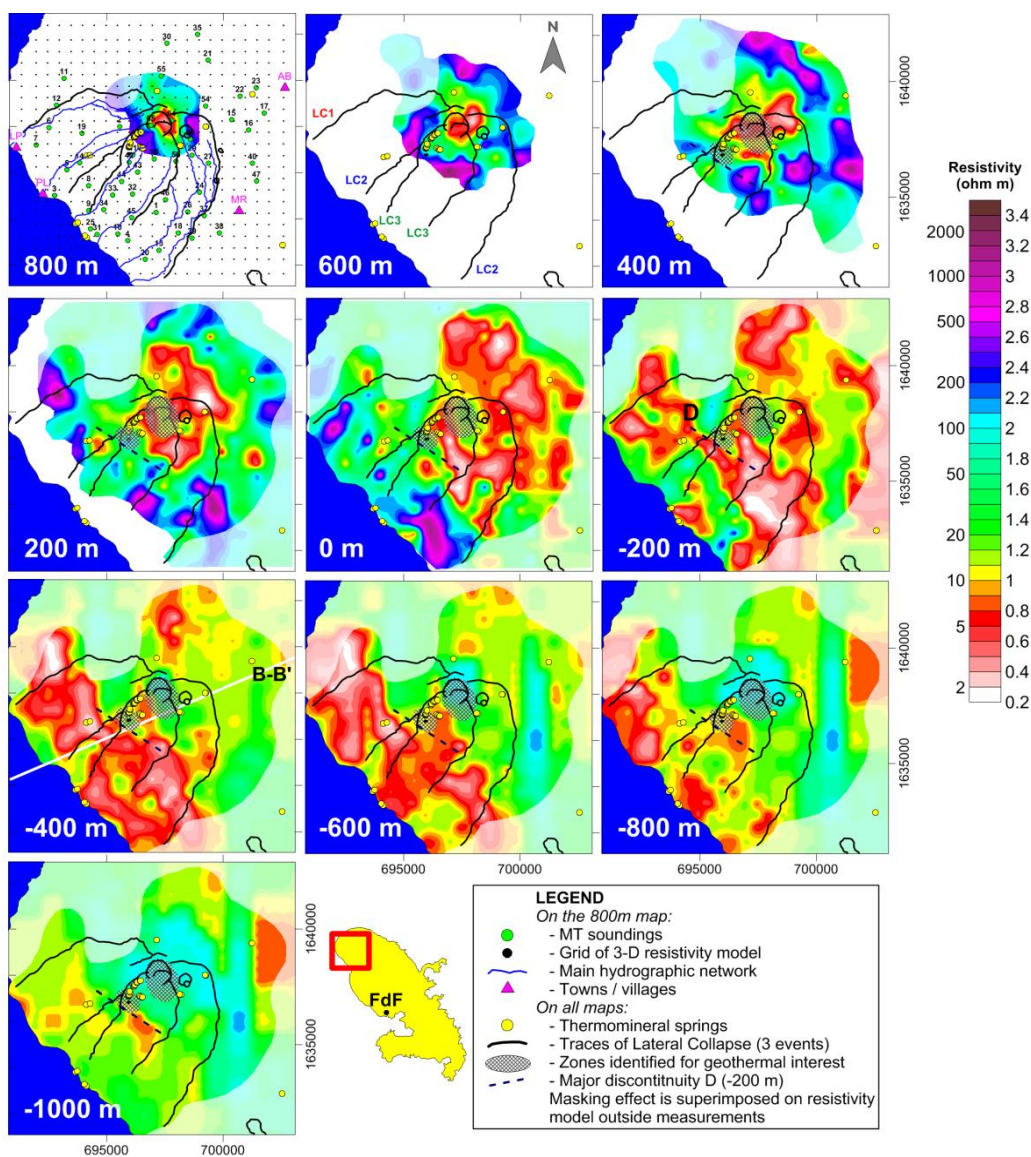

**Figure 5 : Ten horizontal slices extracted from the 3D resistivity model of the Montagne Pelée stratovolcano. Superimposed information is detailed in the legend.**

Coppo et al.

Slightly southwestwards from the summit, we delimited a major discontinuity (D on figure 5, -200 m) based on geophysical data (magnetotelluric and gravity data). Because two different geochemical signatures characterize the springs of Mitan-Picodo Rivières (Cl-Na) and Rivière Chaude (HCO₃-Cl-Na), we infer the potential existence of two separated – at least partly – geothermal reservoirs. This discontinuity could act at depth either as a potential barrier that would favor separated geochemical evolution or as drain for preferential fluid circulation.

Resistivity layers from -400 down to -600 m below sea level point out a highly conductive axis NW-SE orientated, running parallel to the Caribbean sea, and matching gravity lineaments detected on the gradient of the Bouguer anomaly (corrected for a density of 2.0). We may also note on layer -600 b.s.l. that the scars of the latest lateral collapse LC3 (located on figure 5, 600m layer, 9 ka, Le Friant et al., 2003) circumscribes precisely a 5 to 10 Ωm conductive area with higher resistivity on both sides belonging to the NW-SE main axis. This observation is thought to reflect lateral flank collapse processes that excavate one part of the flank of the volcano, including a substantial thickness of the characteristic and general conductive layer found in many volcanic edifices (Descloitres et al., 1997; Nurhasan et al., 2006; Monteiro Santos et al., 2006; Coppo et al., 2009). These catastrophic events deeply affect the stability of the edifice providing potential new preferential pathways for fluids migration and re-formation of the conductive layer which may later acts as cap rock.

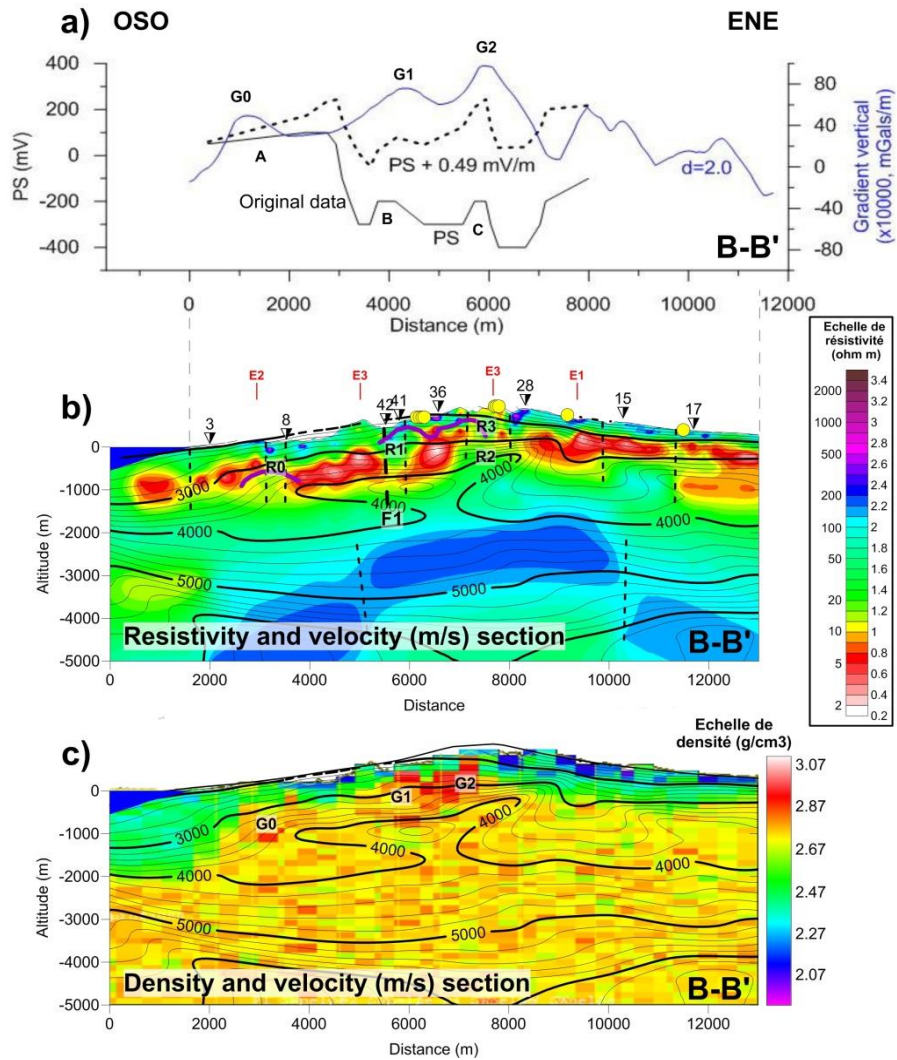

**Figure 6 : Section B-B' (indicated on Figure 5, -400 m map). A) SP original data (Zlotnicki et al., 1998) and corrected for topographical variations, and vertical gradient data (correction density = 2.0). b) Resistivity section extracted from the 3-D resistivity model ( R0, R1, R2 identify more resistive bodies within the main conductive layer) with superimposed velocity model (re-processed from Eschenbrenner et al., 1980). Magenta lines display the top of the denser bodies identified on c). Dashed vertical black line corresponds to the main discontinuity showed on Figure 5. Dashed vertical white lines illustrate discontinuities in the conductive layer. c) Density section extracted from 3-D density model with velocity model. G0, G1, G2 identified denser bodies.**

Although quasi ubiquitous, the conductive layer has a complex morphology that is difficult to connect to a simple a global model with geological and known geothermal manifestations. Particularly, it is difficult to define precisely areas where the conductive layer could act with certainty as cap rock as expected above a well-developed geothermal system. In the classical model of conductive caprock overlying a geothermal, more resistive system (Johnston et al., 1992), the upflow zone is located at the top of the reservoir (Anderson et al., 2000; Usher et al., 2000). In the case of the Montagne Pelée, such typical zone is well highlighted on the -200 m map (Figure 5) where the resistive is surrounded by the conductive layer. The apex of this resistive body is approximately located at altitude 0 m and below the southeastern flank of the summit of the edifice.

To better understand the Montagne Pelée structure, we integrated available SP (Zlotnicki et al., 1998) and refraction seismic (re-processed form Eschenbrenner et al., 1980) data from previous works (Figure 6) along profile B-B'. Some attempts were made to correct the topographical electrokinetic effect on SP in order to better evidence positive anomalies. . Re-interpretation of seismic refraction data show a high velocity anomaly ($Vp = 3000$-$4000$ m/s) between 1000 and 3000 m depth below the Montagne Pelée summit (Figure 6). We thus observe on the SW flank of the volcano, a correlation between, SP positive anomalies (A, B, C), resistive anomalies within the main condcutive layer (R1, R2, R3), dense anomalies of the gravity field (G0, G1 and G2) and a high velocity anomaly at relatively shallow depth as illustrated in figure 6. This configuration of dense and rapid structures may indicate the presence of shallow solidified intrusion possibly related to the present volcano as indicated by the recent lava dome outcropping close to G2 anomaly.

Location of both G1 and G2 dense structures coincident with the uphill scar of the last lateral collapse (LC3 on Figure 5, 600 m), indicate that both structures may be linked. The massive nature of these body and their recent age, both resulting in weak alteration and low clay content can explain the high resistivity anomalies observed and may manage a potential permeability which favours upward fluid flow and associated positive SP anomalies.

This whole set of observations more or less converges towards a general schema of which make it possible the presence of a geothermal system heated by quite superficial and recent intrusion, cap by a general and thick conductive layer which is cut by fault and intrusion on the volcano southwest flank which leave hydrothermal fluid weakly flowing upward and outcrop as the known scarce surface manifestation

## 6. SURFICIAL RESISTIVITY

During the MT survey, several soundings were affected by static shift effect, a physical galvanic effect causing apparent resistivity curves to be shifted relative to each other from an unknown scalar factor, while the phase remain unaffected. The time-independent nature of static shift means that there is no impedance phase associated with the phenomenon (Simpson and Bahr, 2005). Many publications deals about static shift, techniques to recover "true" resistivity (for ex: Bahr, 1988; Sternberg et al., 1988; Jones, 1988) and the need or not to correct it before 3-D inversion (Meju, 1996).

At the beginning of 2013, within the framework of the MarTEM project, a heliborne TEM survey carried out by SkyTEM covered the whole Martinique. With the aim to illustrate the ability of the 3-D inversion scheme (Hautot et al., 2000, 2007) to take into account and reproduce static shift effects.to reproduce surficial resistivity distribution, we compared results of the 3-D resistivity model with the results of the TEM resistivity (1-D stitching). To homogenize TEM distribution, TEM resistivity data are first krigged by layer over the investigated area. The median resistivity of this new grid is then extracted above each MT cells and compared with MT resistivity (Figure 7, right plots). Cells that no contain raw TEM information or that are located too far away from fly lines are removed from the statistics. Results show that 82 % of the remaining cells have a maximum resistivity variation below +/-400% for the Anses d'Arlet prospect, 79 % for the Montagne Pelée.

These results demonstrate that in spite of a fully opposite sampling scale (53 MT soundings and more than 20000 TEM for the Montagne Pelée and 32 MT soundings and more than 6200 TEM), surficial resistivity distribution is quite faithfully reproduced by 3-D inversion.

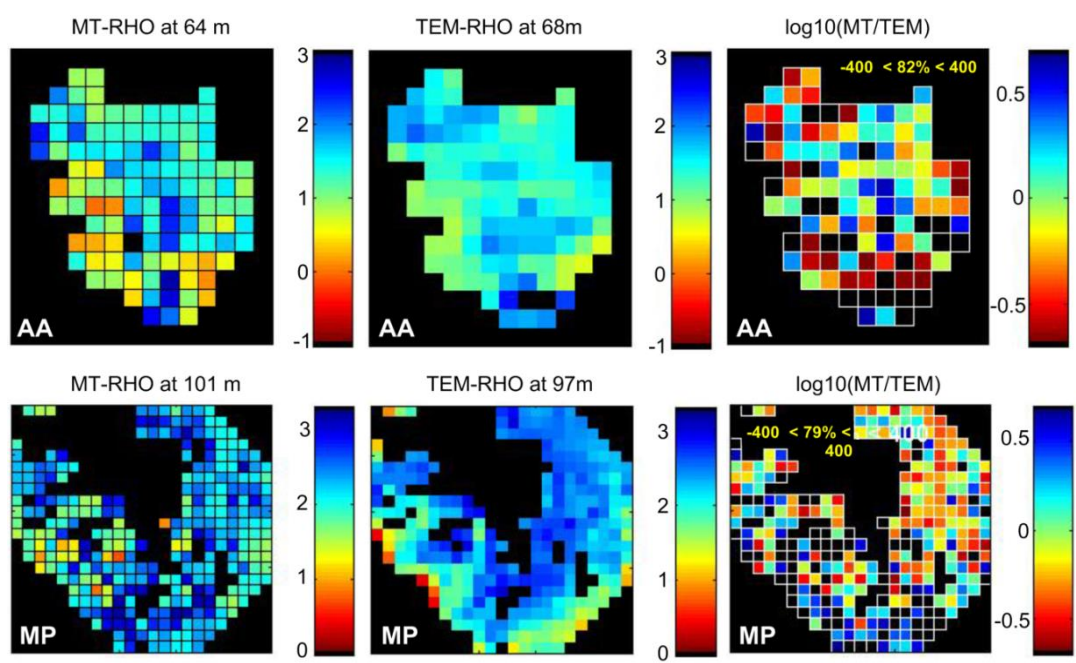

**Figure 7 : MT versus TEM resistivity for superficial layers. Above are represented the data of the Anses d'Arlet (AA) geothermal province and below from the Montagne Pelée (MP) geothermal province. Left plots display resistivity from the 3-D MT models for both sector at the indicated depth. Center plots display resistivity of the TEM data (see text for details). Right plots show variation between the two models (log scale).**

Coppo et al.

# 7. CONCLUSIONS

This paper deals with a part of the geophysical results collected during the last multi-disciplinary geothermal exploration phase of Martinique (2012-2013). We present both the 3-D resistivity models of the Anses d'Arlet and Montagne Pelée geothermal fields resulting of separate 3-D magnetotelluric inversions including sea effect. We point out a potential geothermal resource on each field/area characterized by local specific structures. Complementary geological and geochemical information is provided to help the reader understanding each local context.

On the Anses d'Arlet sector, MT and gravity data reveal a massive dense and resistive body, running parallel to the Caribbean sea and elongated in the NNW-SSE direction. It is interpreted as a N-S trending deep intrusion related to volcanic activity that affected the area during the last My, and could act as heat source of the high temperature geothermal system. A cap rock structure is interpreted at shallow depth. Although many geological and geochemical elements seem to indicate that the geothermal system is currently becoming less active and extensive in space, its southern boundary, which concentrates most of the surface hydrothermal manifestations is proposed to be the best place for further exploration wells.

The geothermal field related to the Montagne Pelée is characterized by a quasi ubiquitous bell-shaped conductive layer and with quite monotonic resistivity distribution making interpretation difficult in terms of geothermal targets. According to geological and geochemical data, we were able to distinguish a potential main discontinuity that could be related to two separated geothermal reservoirs approximately located below the summit of the edifice and southwestwards. Resistivity distribution of the conductive layer seems to be closely connected with edifice evolution and flank collapse events.

Beside these interesting deep structures, we demonstrate, after analyzing the results of the recent heliborne TEM survey covering the whole Martinique island, that surface resistivity distribution obtained from 3-D inversion reproduce faithfully the resistivity distribution observed by TEM. In spite of a very different sampling scale, this comparison illustrates the ability of 3-D MT inversion to take into account and reproduce static shift effects in the sub-surface resistivity distribution.

**ACKNOWLEDGEMENTS**

The authors wish to thanks the FEDER and ADEME institutions, the Conseil Régional and the Syndicat Mixte d'Electricité de la Martinique (SMEM) for funding this project.

**REFERENCES**

Anderson E., Crosby D., Ussher G.: Bulls-eye! - Simple resistivity imaging to reliably locate the geothermal reservoir, *Proceedings*, World Geothermal Congress 2000, Kyushu-Tohoku, Japan, May 28-june 10, 2000, p. 909-914.

Bahr, K.: Interpretation of the magnetotelluric impedance tensor: regional induction and local telluric distortion, Journal of *Geophysics*, **62**, 119-127.

Boudon, G., Villemant B., Le Friant, A., Paterne M., Cortijo E.: Role of large flank-collapse events on magma evolution of volcanoes. Insights from the Lesser Antilles Arc, *Journal of Volcanology and Geothermal Research*, 263, (2013), 224-237.

Briden, J.C., Rex, D.C., Faller, A.M., Tomblin, J.-F.: K–Ar geochronology and palaeomagnetism of volcanic rocks in the Lesser Antilles island arc. *Philosophical Transactions of the Royal Society of London. Series A, Mathematical and Physical Sciences*, **291**, (1979), 485–528.

Caldwell, T. G., Bibby H.M. and Brown C.: The magnetotelluric phase tensor, *Geophysical Journal International,* **158**, (2004), 457-469.

Chave, A. and Thomson D.J.: Bounded influence magnetotelluric response function estimation, *Geophysical Journal International*, **157**, (2004), 988-1006.

Coppo, N., Schnegg, P.-A., Falco, P., Costa, R.: A deep scar in the flank of Tenerife (Canary Islands): Geophysical contribution to tsunami hazard assessment. *Earth Planetary Science Letters*, **282**, (2009), 65-68.

Descloitres, M., Ritz, M., Robineau, B., Courteaud, M.: Electrical structure beneath the eastern collapsed flan of Piton de la Fournaise volcano, Reunion Island: Implications for the quest for groundwater, *Water Resources Research*, **33 (1)**, (1997), 13-19.

Eschenbrenner S., Dorel J. and Viode J.P. : - Coupes sismiques des structures superficielles dans les Petites Antilles – II : Martinique. Pageoph., Birkhäuser Verlag, Basel, **vol. 118**, (1980), p. 807-822.

Gadalia, A., J.M. Baltassat, V. Bouchot, S. Caritg, N. Coppo, F. Gal, J.F. Girard, A. Gutierrez, T. Jacob, G. Martelet, S. Rad, A.L. Tailame, H. Traineau, B. Vittecoq, P. Wawrzyniak, C. Zammit : Compléments d'exploration géothermique en Martinique : conclusions et recommandations pour les zones de la Montagne Pelée, des Anses d'Arlet, des Pitons du Carbet et du Lamentin, 2014), Rapport final BRGM/RP-FR, 227 p, 75 fig., 7 tabl., 4 ann., 1 CD.

Germa, A., Quidelleur, X., Labanieh, S., Lahitte, P., Chauvel, C.: The eruptive history of Morne Jacob volcano (Martinique Island, French West Indies): geochronology, geomorphology and geochemistry of the earliest volcanism in the recent Lesser Antilles arc, *Journal of Volcanology and Geothermal Research*, **198**, (2010), 297–310.

Germa, A., Quidelleur, X., Lahitte, P., Labanieh, S., Chauvel, C.: The K–Ar Cassignol–Gillot technique applied to western Martinique lavas: a record of the evolution of the recent Lesser Antilles island arc activity from 2 Ma to Mount Pelée volcanism, *Quaternary Geochronology*, **6**, (2011), 341–355.

Germa, A., Quidelleur, X., Labanieh, S., Chauvel, C. and Lahitte, P.: The volcanic evolution of Martinique Island : Insights from K-Ar dating into the Lesser Antilles arc migration since the Oligocene, *Journal of Volcanology and Geothermal Research*, **208**, (2011), 122-135.

Hautot, S., Tarits, P., Whaler, K., Le Gall, B., Tiercelin, J.J, and Le Turdu, C.: The deep structure of the Baringo Rift basin (central Kenya) from 3-D magneto-telluric imaging: Implications for rift evolution, *Journal of Geophysical Research*, **105**, (2000), 23493-23518.

Hautot, S., R. Single, J. Watson, N. Harrop, D. A. Jerram, P. Tarits, and K. A. Whaler.: 3-D magnetotelluric inversion and model validation with gravity data for the investigation of large igneous provinces, *Geophysical Journal International*, **170(3)**, (2007), 1418-1430.

Johnston, J. M., Pellerin, L., Hohmann, G.W.: Evaluation of electromagnetic methods for geothermal reservoir detection, *Geothermal ressources transactions*, vol. 16, (1992).

Jones, A.G.: Static-shift of magnetotelluric data and its removal in a sedimentary basin environment, *Geophysics*, **53**, (1988), 967-978.

Labanieh, S.: Géochimie de l'île de la Martinique aux Petites Antilles. PhD Thesis, Université Joseph Fourier, Grenoble, (2009), 278 pp. Available on line at http://tel.archives-ouvertes.fr/tel-00467762/fr/. In French, English abstract.

Le Friant A., Boudon G., Deplus C., Villemant B.: Large-scale flank collapse events during the activity of Mount Pelée, Martinique, Lesser Antilles, *Journal of Geophysical Research*, **108(B1)**, (2003), 2055, doi:10.1029/2001JB001624.

Meju, M.A.: Joint inversion of TEM and distorted MT soundings: some effective practical consideration, *Geophysics*, **61**, (1996), 56-65.

Monteiro Santos, F.A., Trota, A., Soares, A., Luzio, R., Lourenço, N., Matos, L., Almeida, E., Gaspar, J.L., Miranda, J.M.: An audio-magnetotelluric investigation in Terceira Island (Azores). *Journal of Applied Geophysics*, **59(4)**, (2006), 314-323.

Nagle, F., Stipp, J.J., Fisher, D.E.: K–Ar geochronology of the Limestone Caribbees and Martinique, Lesser Antilles, West Indies, *Earth and Planetary Science Letters*, **29**, (1976), 401–412.

Nurhasan, Ogawa, Y., Ujihara, N., Tank, B., Honkura, Y., Onizawa, S., Mori, T., Makino, M.: Two electrical conductors beneath Kusatsu-Shirane volcano, Japan, imaged by audiomagnetotellurics, and their implications for the hydrothermal system, *Earth Planets Space*, **58**, (2006), 1053-1059.

Simpson, F. and K. Bahr, Practical Magnetotellurics. Cambridge, (2005), Cambridge University Press.

Smith, A.L., Roobol, M.J.: Mt Pelee, Martinique: a study of an active island-arc volcano, *The Geological Society of America Memoir*, **175**, (1990), 105 pp.

Sternberg, B. K., Washburne, J.C. and Pellerin L.: Correction for the static shift in magnetotellurics using transient electromagnetic soundings, *Geophysics*, **53(11)**, (1988), 1459-1468.

Traineau H., Bouchot V., Caritg S., Gadalia A.: Compléments d'exploration géothermique en Martinique : volet géologie, Rapport intermédiaire, (2013), BRGM/RP-62349-FR, 153 p., 69 fig., 9 tabl., 3 ann.

Quidelleur X., Hildenbrand A., Samper A.: Causal link between Quaternary paleoclimatic changes and volcanic islands evolution, *Geophysical Research Letters, 35* **(2)**, (2008), L02303, doi:10.1029/2007GL031849.

Traineau H., Bouchot V., Caritg S., Gadalia A.: Compléments d'exploration géothermique en Martinique : volet géologie, (2013), Rapport intermédiaire, BRGM/RP-62349-FR, 153 p., 69 fig., 9 tabl., 3 ann.

Vozoff, K. (1991). The magnetotelluric method, Electromagnetic Methods in Geophysics. M. N. Nabighian. Tulsa, Oklahoma,USA, Society of Exploration Geophysicists, **2**, (1991), 641-711.

Ussher, G., Harvey, C., Johnstone, R.: Understanding the resistivities observed in geothermal systems, *Proceedings*, World Geothermal Congress 2000, Kyushu-Tohoku, Japan, May 28-june 10, 2000, p. 1915-1920.

Wawrzyniak, P. Sailhac P., and Marquis G.: Robust error on magnetotelluric impedance estimates, *Geophysical Prospecting*, **61**, 533-546, DOI: 10.1111/j.1365-2478.2012.01094.x.

Westercamp, D.: Diversité, contrôle structural et origines du volcanisme récent dans l'arc insulaire des Petites Antilles. *Bulletin du BRGM, deuxième série section IV (*no **3/4**), (1979), 211–226.

Westercamp D. and Traineau H.: Carte géologique de la Montagne Pelée à 1/20 000 (département de la Martinique) et notice explicative simplifiée, (1983), Éd. BRGM.

Westercamp, D., Andreieff, P., Bouysse, P., Cottez, S., Battistini, R. : Martinique; carte géologique à 1/50 000. In: BRGM (Ed.), (1989), 246 pp.